\documentclass[traditabstract]{aa} 
\usepackage{graphicx}
\usepackage{txfonts}
\usepackage{natbib,hyperref}
 \usepackage[modulo,switch]{lineno}
\usepackage{multirow}
\usepackage{longtable}

\bibpunct{(}{)}{;}{a}{}{,} 

\usepackage{graphicx}
 \usepackage{float}
\usepackage{txfonts}

\begin{document}

   \title{Photometric and spectroscopic variability of the B5IIIe star HD\,171219\thanks{The CoRoT  space mission was developed and operated by the French space agency CNES, with participation of ESA's RSSD and Science Programmes, Austria, Belgium, Brazil, Germany, and Spain. This work is partially based on observations made with the 3.6-m telescope at La Silla Observatory under the ESO Large Programme LP185.D-0056.}}

   \titlerunning{Photometric and spectroscopic variability of the B5IIIe star HD\,171219}

   \authorrunning{Andrade et al.}


   \author{L. Andrade\inst{1},
          E. Janot-Pacheco\inst{2},
          M. Emilio\inst{1,3},
          Y. Fr\'emat\inst{4},
          C. Neiner\inst{5},
          E. Poretti\inst{6},
          P. Mathias\inst{7},
          M. Rainer\inst{6},
          J.C. Su\'arez\inst{8,9},
          K. Uytterhoeven\inst{10},
          M. Briquet\inst{5,11},
          P.D. Diago\inst{12},
          J. Fabregat\inst{13}
          \and
          J. Guti\'errez-Soto\inst{12}
          }

   \institute{Observat\'orio Astron\^omico/DEGEO, Universidade Estadual de Ponta Grossa,
              Av. Carlos Cavalcanti, 4748, 84030-900, Ponta Grossa, PR, Brazil, \email{laerteandrade@uepg.br}
         \and
             Universidade de S\~ao Paulo/IAG-USP, rua do Mat\~ao, 1226 - Cidade Universit\'aria, 05508-900, S\~ao Paulo, SP, Brazil
         \and
             Institute for Astronomy, University of Hawaii, Honolulu, HI 96822, USA
         \and
             Royal Observatory of Belgium, 3 avenue circulaire, B1180 Brussels, Belgium
         \and
             LESIA, Observatoire de Paris, PSL Research University, CNRS, Sorbonne Universit\'es, UPMC Univ. Paris 06, Univ. Paris Diderot, Sorbonne Paris Cit\'e, 5 place Jules Janssen, 92195 Meudon, France
         \and
             INAF–Osservatorio Astronomico di Brera, via E. Bianchi 46, 23807 Merate (LC), Italy
         \and
             Institut de Recherche en Astrophysique et Plan\'etologie, CNRS, 14 avenue Edouard Belin,
Universit\'e de Toulouse, UPS-OMP, IRAP, 31400 Toulouse, France
         \and
             Universidad de Granada, Dept. F\'isica Te\'orica y del Cosmos, 18071, Granada, Spain
         \and
             Instituto de Astrof\'isica de Andaluc\'ia (CSIC), Glorieta de la Astronom\'ia s/n, 18008 Granada, Spain
         \and
             Instituto de Astrof\'isica de Canarias, E-38205 La Laguna, Tenerife, Spain
         \and
             Institut d’Astrophysique et de G\'eophysique, Universit\'e de Li\`ege, Quartier Agora, All\'ee du 6 ao\^ut 19C, 4000 Li\`ege, Belgium
         \and
	     Depto. Did\'actica de la Matem\'atica, Universidad de Valencia, Avda. Tarongers, 4, 46022 Valencia, Spain
         \and
             Observatorio Astron\'omico, Universidad de Valencia, 46100 Burjassot, Spain
             }

   \date{Received: June 23, 2016; accepted: March 24, 2017}


\abstract{We analyzed the star HD\,171219, one of the relatively bright Be stars observed in the seismo field of the CoRoT satellite, in order to determine its physical and pulsation characteristics. Classical Be stars are main-sequence objects of mainly B-type, whose spectra show, or had shown at some epoch, Balmer lines in emission and an infrared excess. Both characteristics are attributed to an equatorially concentrated circumstellar disk fed by non-periodic mass-loss episodes (outbursts). Be stars often show nonradial pulsation gravity modes and, as more recently discovered, stochastically excited oscillations. Applying the CLEANEST algorithm to the high-cadence and highly photometrically precise measurements of the HD\,171219 light curve led us to perform an unprecedented detailed analysis of its nonradial pulsations. Tens of frequencies have been detected in the object compatible with nonradial g-modes. Additional high-resolution ground-based spectroscopic observations were obtained at La Silla (HARPS) and Haute Provence (SOPHIE) observatories during the month preceding CoRoT observations. Additional information was obtained from low-resolution spectra from the BeSS database. From spectral line fitting we determined physical parameters of the star, which is seen equator-on ($i=90^{\circ}$). We also found in the ground data the same frequencies as in CoRoT data. Additionally, we analyzed the circumstellar activity through the traditional method of violet to red emission H$\alpha$ line variation. A quintuplet was identified at approximately $1.113$ c\,d$^{-1}$ (12.88 $\mu$Hz) with a separation of $0.017$ c\,d$^{-1}$ that can be attributed to a pulsation degree $\ell\sim2$. The light curve shows six small- to medium-scale outbursts during the CoRoT observations. The intensity of the main frequencies varies after each outburst, suggesting a possible correlation between the nonradial pulsations regime and the feeding of the envelope.}


   \maketitle

%

\section{Introduction}

Classical Be stars are main-sequence objects of mainly B-type, whose spectrum show, or has shown at some epoch, Balmer lines in emission and an infrared excess. Both characteristics are attributed to an equatorially concentrated circumstellar disk produced by sporadic mass ejections occurring during light brightening episodes, called outbursts. The ejection episodes can be explained by the rapid rotation of these objects, but since most Be stars rotate below their critical rotation velocities \citep{fre05, cra05, yud01} an additional explanation is needed. Because nonradial pulsations (NRP) in the Be star \object{$\mu$ Cen} \citep{riv98} correlated with mass ejection episodes, NRP could provide the additional energy required to trigger the mass ejection episodes, converting a high V$\sin i$ B star into a Be star. \citet{hua09} found for the hybrid pulsating Be \object{HD 49330}, from CoRoT and ground-based spectroscopic data, an increase in amplitude for g-modes and a decrease for p-modes just before an outburst and a reverse behavior after the outburst. \citet{nei12} discovered stochastically-excited gravito-inertial modes in the Be star, \object{HD 51452}, enhanced by rapid rotation. They found that the mode amplitudes were linked to the presence of very-small-scale outbursts. This and \citet{hua09} led them to propose a scenario to explain Be outbursts by the transport of angular momentum by waves \citep{nei13, nei14}. Despite the empirical evidence, the physical process responsible for the outbursts and mass loss in Be stars is still poorly understood. Furthermore, the effect of an outburst on NRP mode production by means of change in stellar structure is worth investigating \citep[see][for a comprehensive review on Be stars]{riv13}.

\citet{dzi93} showed that high-order g-modes driven by the $\kappa$ mechanism are unstable in the region of the HR diagram occupied by Be stars. The predicted radial velocity amplitude caused by an $\ell$ = 3 g-mode is $\sim$ 20 km\,s$^{-1}$ for a 4 $M_{\sun}$ star for pulsation periods in the range $\sim$ 0.5 – 1.1 d. For more massive and hotter stars such as \object{HD 171219}, unstable modes are predicted with periods of 0.5 – 2 d.  \citet{tow05} found that mixed NRP modes (hybrid between Rossby and Poincaré modes) are unstable for B stars. Inertial-frame typical periods for Be stars occur in the range $\sim$ 0.2 – 2.0 d.  

Pulsation analysis of a star requires the measurement of frequencies with both high photometric precision and high time-frequency resolution in addition to being free of one-day aliases, typical of ground-based observations. Only space missions provide both conditions. Stellar seismology made a great leap forward thanks to MOST, CoRoT, and Kepler satellites.

The present paper reports the analysis of the Be star HD\,171219. The observations with CoRoT and ground-based spectroscopy are described in Sect.~\ref{sect:observations}. The star parameters, its light intensity behavior, and circumstellar variations are presented in Sect.\ref{sect:thestar}. We discuss the pulsation analysis  in Sect.~\ref{sect:cleanest} and present our conclusions in Sect.~\ref{sect:conclusions}.

\section{Observations} \label{sect:observations}

\subsection{CoRoT photometric observations}

The main light curve of HD\,171219 was obtained with the seismology core program of the CoRoT (Convection, Rotation and planetary Transits) space observatory \citep{bag16}, which studied bright pulsating stars $(6<V<9)$ with cadences of 32 and 512 seconds. For detailed information on the CoRoT satellite, see \citet{auv09} and \citet{sam07}.

\begin{figure}
 \resizebox{\hsize}{!}{\includegraphics[angle=90]{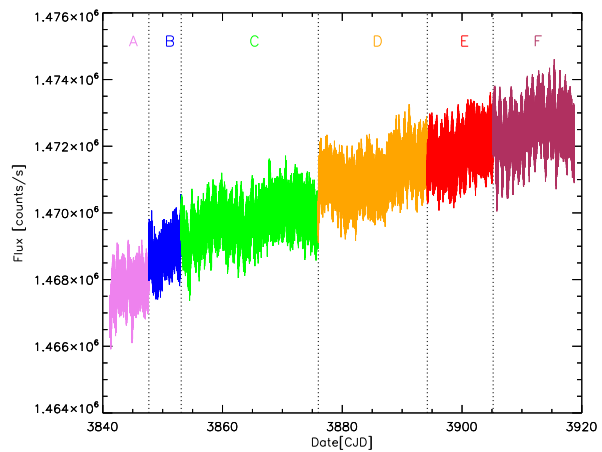}}
 \caption{CoRoT light curve of the Be star HD\,171219 in the satellite reference frame (CoRoT Julian Date, CJD = HJD $- 2455000$ d). The colored regions (A) through (F) are divided by the occurrences of a medium and minor outbursts.}
 \label{lightcurve}
 \end{figure}

Data on HD\,171219 were taken during 77.56 d of the LRc06 long run from July 08th 2010 to September 24th 2010 with a cadence of 32 seconds. The light curve is shown in Fig.~\ref{lightcurve}. It contains 209\,418 photometric observations. Five small-scale outbursts were observed in the light curve during the CoroT observations. For the purpose of the pulsation analysis we then divided the light curve into six regions as shown in Fig.~\ref{lightcurve} (see section~\ref{fourieranalysis}).

\subsection{Ground-based spectroscopic observations}

Spectroscopic observations of HD\,171219 were obtained with HARPS (3.6m telescope, ESO, La Silla, Chile) and SOPHIE (1.93m telescope, Haute Provence Observatory, OHP, France) spectrographs as a part of the ground-based campaign on the CoRoT targets (see Table~\ref{table:1}). We also used low-resolution observations from the BeSS database \citep{nei11} from about the same epoch.

\begin{figure}
 \resizebox{\hsize}{!}{\includegraphics[angle=90]{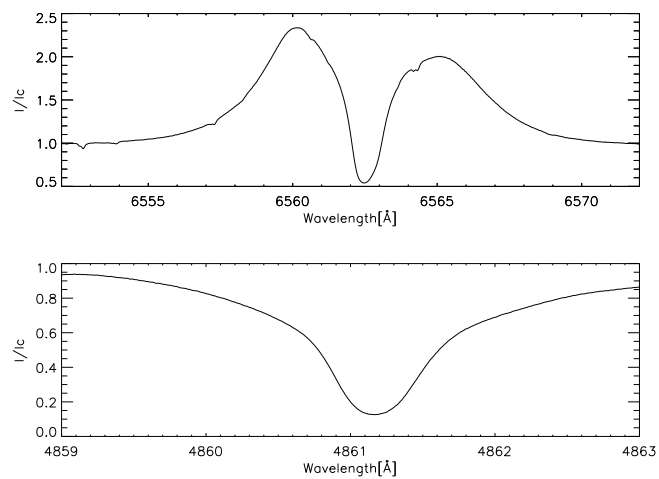}}
 \caption{Average HARPS spectrum of HD\,171219 centered around the H$\alpha$ line (top panel) and the H$\beta$ line (bottom panel).}
\label{halpha}
\end{figure}

HARPS was used in the high-efficiency EGGS mode $R$=80,000 with a fast readout \citep{may03}. From June 14 to  July 2, 2010, 41 spectra were collected. The average exposure time was 1200s and the signal-to-noise ratio (S/N) ranges from 97 to 236. HARPS spectra are available in the SISMA database \citep{rai16}. Strong fringes in the reduced data were carefully corrected, as detailed in \citet{nei12}. Figure~\ref{halpha} shows the average HARPS spectrum around two of the Balmer lines. The H$\alpha$ line profile depicts the status of emission activity during observations.

SOPHIE was used in the high efficiency mode (HE) yielding a resolving power of $R$=40,000. Twelve spectra were obtained in late June, 2010 with an average exposure time of about 900s. The resulting S/N ratio varies from 90 to 213. Usual bias, flatfield and wavelength calibration was performed using the SOPHIE pipeline.

We also retrieved reduced spectra in the H$\alpha$ line region obtained by amateur astronomers in June and July 2010 from the BeSS database. The data were obtained with telescopes ranging from 21 to 30 cm in diameter, and a resolving power of $R \sim$ 15,000.

\begin{table}
\caption{Spectroscopy observations of HD\,171219 obtained close to the CoRoT photometric observations}
\label{table:1}
\centering
\begin{tabular}{ l l l l }
\hline\hline
DATE & Instrument & \# of Spectra & Covered Time[hours]\\
\hline
2010.06.14 & HARPS & 02 & 01\\
2010.06.16 & HARPS & 07 & 07\\
2010.06.18 & HARPS & 03 & 03\\
2010.06.21 & HARPS & 06 & 07\\
2010.06.22 & HARPS & 05 & 06\\
2010.06.22 & SOPHIE & 01 & \\
2010.06.23 & SOPHIE & 02 & 01\\
2010.06.24 & SOPHIE & 02 & 01\\
2010.06.25 & SOPHIE & 02 & 01\\
2010.06.26 & SOPHIE & 02 & 01\\
2010.06.27 & SOPHIE & 02 & 01\\
2010.06.28 & SOPHIE & 01 & \\
2010.07.01 & HARPS & 11 & 07\\
2010.07.02 & HARPS & 07 & 07\\
\hline
\end{tabular}
\end{table}

\begin{figure}
\resizebox{0.9\hsize}{!}{\includegraphics{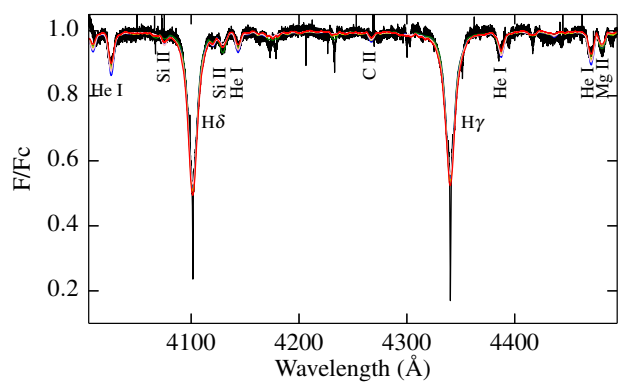}}
\caption{Average SOPHIE blue spectrum of HD\,171219 (black points). Red line is our best fit used to derive the stellar parameters (Table~\ref{table:2}) and green line is the solution obtained previously with only one FEROS spectrum \citep{fre06}. We highlight the sharp bottom shape of Balmer shell lines. Main spectral lines used to derive the apparent parameters are identified.}
\label{fit_complete}
\end{figure}

\begin{figure}
\resizebox{0.9\hsize}{!}{\includegraphics{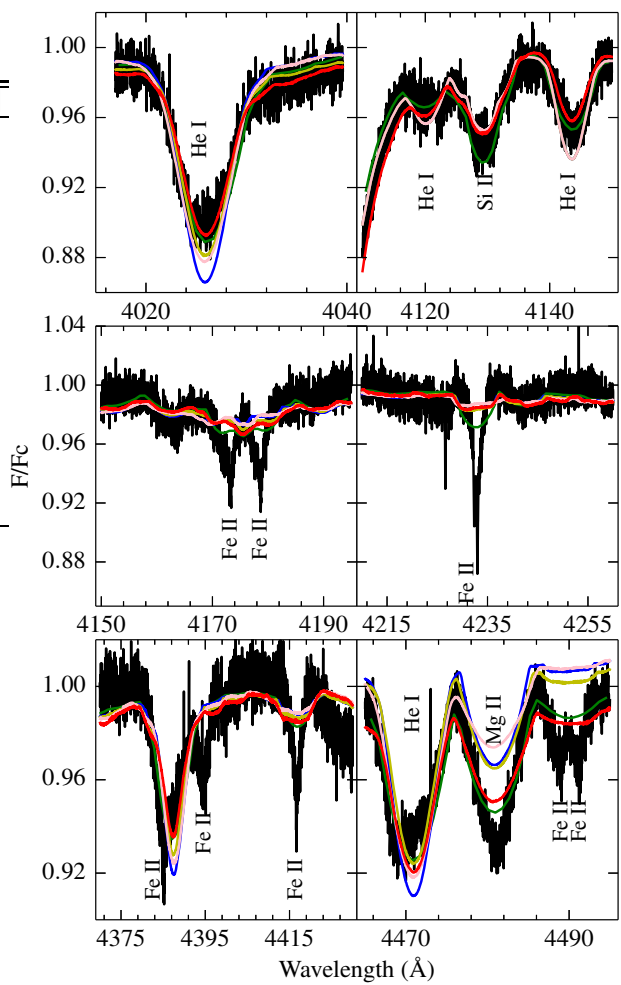}}
 \caption{The same as Fig.~\ref{fit_complete} for some spectral regions. Predicted photospheric Fe II lines (red and green curves) are top flattened as expected from line transitions formed at the stellar equator. In the observations, they are dominated by the shell features.}
 \label{fit_parts}
\end{figure}

\section{The Be star HD\,171219} \label{sect:thestar}

\subsection{Stellar parameters}

HD\,171219 is a classical  $V$= 7.65 mag Be star. First values of the fundamental parameters of the star were given in \citet{fre06}. This determination was based on one FEROS spectrum with S/N(5500~\AA) $\sim$ 130. Apparent parameters, not considering gravitational darkening, were derived by fitting synthetic spectra with plane-parallel model atmospheres (1D spectra) to the FEROS data. A table of corrections was then used to correct the apparent parameters values for gravitational darkening and estimate the fundamental parameters of the non-rotating stellar counterpart as well as the "true" V$\sin i$. Currently, 53 good quality spectra obtained with the SOPHIE and HARPS spectrographs are available. We decided to rederive the apparent parameters by directly comparing 2D spectra that account for first-order rapid-rotation effects (stellar flattening and gravitational darkening) to observations made of a combination of the best available spectra. In the procedure, the spectra were computed with the FASTROT routine, using the same atomic data and atmosphere models as in \citet{fre06}.

The presence of quite stable, deep, narrow shell-like absorptions superimposed to the broader photospheric lines over the period of observations makes any fundamental parameter determination very difficult. We therefore started using 1D spectra to fit the wings of the helium and hydrogen lines present in the $\lambda$ 4000-4500 wavelength range. Since the $\lambda$ 4026 \ion{He}{i} line did not seem to be affected by shell lines, we added the complete profile in the subsequent fits and rederived the apparent parameters and V$\sin i$ which were found to be equivalent to those given in \citet{fre06}, as can be seen in the green lines in Figs.~\ref{fit_complete} and~\ref{fit_parts}. In the next step, a grid of 2D spectra was computed for $\Omega / \Omega_{\rm c}$ = 0.90, 0.95, and 0.99, $i$ = 70 - 90$^{\rm o}$ ($\Delta$i = 5$^{\rm o}$, $T_{\rm eff}$ = 12000 - 18000 K ($\Delta T_{\rm eff}$ = 100 K), log(g) = 3.30 - 4.20 ($\Delta$log g = 0.1). Lower inclinations $i$ and rotation rates were not considered because they were not providing enough broadening to account for the extended helium line wings.

Each synthetic spectrum was compared individually to the observations and the closest match was obtained for the parameters given in Table~\ref{table:2}. The best spectrum is shown in red in Figs.~\ref{fit_complete} and~\ref{fit_parts} and compared to the 1D spectrum (green line). As the figures show, most shell lines are attributable to \ion{Fe}{ii} and appear top flattened in the photospheric 2D synthetic spectrum, as we expect for transitions formed at the equator in the cooler parts of the star's surface. The determined stellar parameters are compatible with a rapidly rotating B5 III star.

We find that $i=90^{\circ}$, that is, the star is seen equator-on. Moreover $\Omega/\Omega_c$ is $0.99$, and thus the star rotates close to critical rotation.

\begin{table}
\caption{Stellar parameters of HD\,171219. Superscript "nrcp" indicates the derivation for a non-rotating counterpart of the star \citep{fre05}. Subscript "true" indicates the apparent V$\sin i$ value corrected for gravitational darkening effects.}
\label{table:2}
\centering
\begin{tabular}{ l l }
\hline\hline
Parameter & Value\\
\hline
$T_{\rm eff}^{\rm nrcp}$ (K) & 15000\\
$\log g^{\rm nrcp}$ & 3.60\\
$i$ & $90^{\circ}$ \\
V$\sin i_{\rm true}$ (km\,s$^{-1}$)  & 345\\
$\Omega/\Omega_{\rm c}$ & 0.99\\
$f_{\rm rot}$ $(c d^{-1})$ & 0.833 \\
$R_{\rm equ}$ (R$_{\odot}$) & 8.18 \\
$R_{\rm equ}/R_{\rm pole}$ &  1.46\\
$T_{\rm eff}^{\rm pole}$ (K) & 16735\\
$T_{\rm eff}^{\rm equ}$ (K) & 7632\\
$\log g^{\rm pole}$ & 3.68\\
$\log g^{\rm equ}$ &2.30\\
$V_{\rm crit}$ (km\,s$^{-1}$) & 356\\
\hline
\end{tabular}
\end{table}

\subsection{Circumstellar variations and occurrence of minor outbursts}

Be stars undergo aperiodic mass ejection episodes that feed their circumstellar disks. These so-called outbursts produce brightness variations and changes in the emission line profiles. Be stars are known to present short and long-term variations of the violet (V) and red (R) components of emission lines (V/R variations). The long-term variations (5 to 10 years) can be attributed to one-armed oscillations in the Be circumstellar envelope \citep{oka97}. Short-term variations were first thought to be caused by rotating circumstellar structures but space-based observations give increasing evidence that they are caused by pulsations \citep{riv13}. There are also cases where the V/R variability was shown to be phase-locked to orbital period in binaries \citep{zha13}.

Close examination of the HD\,171219 CoRoT light curve shows the occurrence of a medium-sized outburst around CJD (HJD $- 2455000$) = 3876 and four minor outbursts close to CJD = 3847, 3853, 3894 and 3905 (see Fig.~\ref{lightcurve}). Some of these outbursts could actually be instrumental jumps in the light curve. Such jumps have been observed in other (non-Be) stars. The only way to be sure that they are outbursts is to check light curves of other stars observed at the same time in the same CCD. If they also show a jump at the same time, then it is an instrumental jump. Only HD\,171219 showed a jump in our verification, proving that it is in fact a real outburst. The relative intensity of pulsation frequencies found in the star change between the outbursts (see section~\ref{fourieranalysis}).

\begin{figure}
 \resizebox{\hsize}{!}{\includegraphics{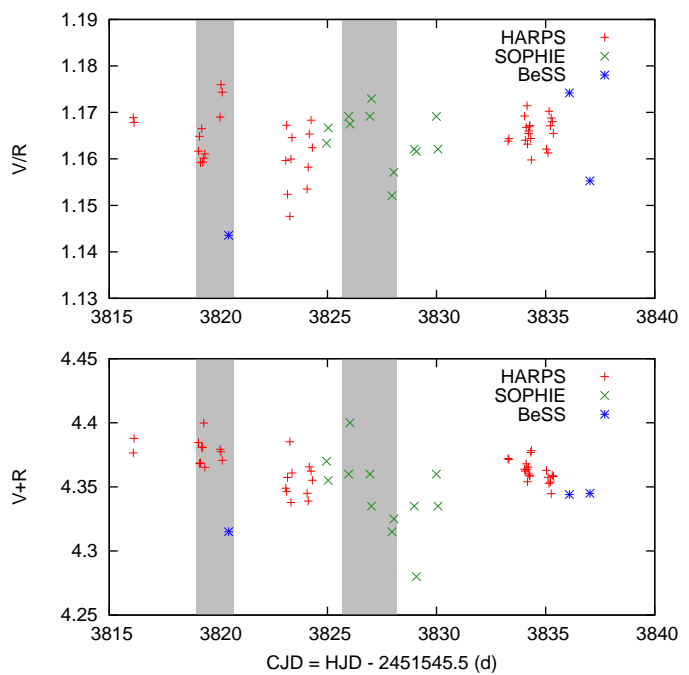}}
 \caption{Violet over red (V/R) emission ratio (top panel) and sum of
the V and R emission (bottom panel) measured in the emission peaks of the H$\alpha$
line observed with HARPS, SOPHIE, and additional spectra from the BeSS
database. The hashed regions indicate the occurrence of minor
outbursts.}
 \label{VRrelations}
 \end{figure}

We measured the relative intensity of the violet to red components of the H$\alpha$ emission line from HARPS and SOPHIE observations, namely $V/R = [V - V_c]/[R-R_c]$ \citep{car03}. Three measurements from the BeSS database \citep{nei11} were also included (see Fig.~\ref{VRrelations}).

The $V/R$ variations indicate two epochs of sudden change in $V/R$, centered around CJD 3818 and 3828. In both cases there is a definite increase in the $V/R$ ratio and in the $V+R$ relation, which can be attributed to minor outbursts. The time interval between these episodes is compatible with the outburst occurrence (10 to 20 d) seen in the CoRoT observations that began the following month.

\section{Pulsation analysis} \label{sect:cleanest}

Fourier analysis of photometric and spectroscopic data was performed with the CLEANEST algorithm~\citep{fos95, fos96}. It removes most of the alias patterns in time-series analysis as discussed previously in \citet{emi10}.

\subsection{Time-frequency analysis of CoRoT data}

\subsubsection{Time-Frequency analysis of the whole light curve}

We assume a precision $\Delta \nu$ on the frequency determination equal to the inverse of the duration of the observing run, that is, $\Delta \nu = 1/77.56$ d = $0.013$ c\,d$^{-1}$. Periods shorter than about one minute ($\nu = 1440$ c\,d$^{-1}$) and longer than the total duration of the observing run of 77.56 d ($\nu = 0.013$ c\,d$^{-1}$) could not be detected.

\begin{figure}
 \resizebox{\hsize}{!}{\includegraphics{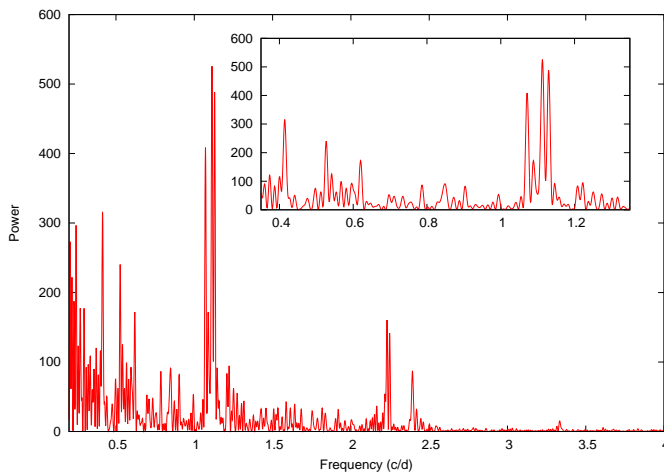}}
 \caption{Power spectrum of the frequencies extracted from the full CoRoT
data (Fig.~\ref{lightcurve}) using CLEANEST. Inset: Zoom of the region containing the
most relevant powerful frequencies (0.4 - 1.3 c/d).}
 \label{cleanest}
 \end{figure}

We applied the  CLEANEST algorithm to look for pulsation frequencies on the HD\,171219 light curve.  In  Fig.~\ref{cleanest} we show the CLEANEST spectrum of the reconstructed time series. Table~\ref{table:3} shows a list of the main frequencies found. Frequencies that correspond to the star rotation period, $0.849$ c\,d$^{-1}$ (compatible with the stellar parameters in Table~\ref{table:2} within our frequency precision) and a half of this period appear in the CLEANEST power spectrum. 

The ensemble of frequencies ($f_{1}=1.072$ c\,d$^{-1}$, $f_{2}=1.089$ c\,d$^{-1}$, $f_{3}=1.113$ c\,d$^{-1}$, $f_{4}=1.130$ c\,d$^{-1}$ and $f_{5}=1.146$ c\,d$^{-1}$) seems to be a g-mode quintuplet. These frequencies are above the significance limit of 99.9\%. They are equidistant with $f \simeq 0.017$ c\,d$^{-1}$.  The distance between $f_2$ and $f_3$ is $0.024$ c\,d$^{-1}$, compatible with $0.017$ c\,d$^{-1}$ within the precision in our frequency determination from the light curve ($0.013$ c\,d$^{-1}$). Deviations from equidistancy are expected for multiplets of fast rotating stars \citep{pro81}. Differential rotation can also alter the frequencies of the modes.

First harmonics of some of these frequencies can be seen at about $2.2$ c\,d$^{-1}$.  There is also a group of less regularly spaced, relatively strong frequencies around $0.4$ -- $0.6$ c\,d$^{-1}$.  As will be seen in the following section, each one of the quintuplet frequencies found in the whole time-frequency analysis appear with different power at different times of the photometric observations.

\begin{table}
\caption{The most powerful frequencies detected in CoRoT data with CLEANEST}
\label{table:3}
\centering
\begin{tabular}{ l c r r }
\hline\hline
\multicolumn{3}{c}{Frequency} \\
\cline{1-3}
 Remarks        & c\,d$^{-1}$   & $\mu$Hz            & Power \\
\hline
                & 0.401         & 4.64               & 116 \\
$f_{\rm rot}$/2 & 0.414         & 4.79               & 315  \\
                & 0.527         & 6.10               & 240 \\
$f_2$/2         & 0.542         & 6.27               & 125 \\
$f_4$/2         & 0.568         & 6.57               & 99  \\
                & 0.596         & 6.90               & 93  \\
                & 0.620         & 7.18               & 172 \\
                & 0.787         & 9.11               & 86  \\
$f_{\rm rot}$   & 0.849         & 9.83               & 91  \\
                & 0.904         & 10.46              & 82  \\
$f_1$           & 1.072         & 12.41              & 409 \\
$f_2$           & 1.089         & 12.60              & 172 \\
$f_3$           & 1.113         & 12.88              & 526 \\
$f_4$           & 1.130         & 13.08              & 488 \\
$f_5$           & 1.146         & 13.26              & 92  \\
                & 1.209         & 13.99              & 83  \\
                & 1.223         & 14.16              & 95  \\
2*$f_3$         & 2.231         & 25.82              & 160 \\
2*$f_4$         & 2.248         & 26.02              & 141 \\
                & 2.393         & 27.70              & 86  \\
\hline
\end{tabular}
\end{table}

 \begin{figure}
 \resizebox{\hsize}{!}{\includegraphics[angle=90]{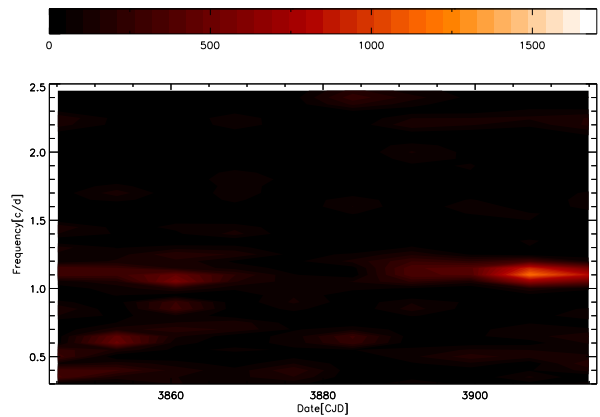}}
 \caption{Periodogram of the results of the time-frequency analysis of
the CoRoT light curve (Fig.~\ref{lightcurve}) of HD\,171219 showing the variation in
amplitude, including disappearance and reappearance, of the main
frequencies during the CoRoT run.}
 \label{fig03}
 \end{figure}

\subsubsection{Fourier analysis between outbursts} \label{fourieranalysis}

The five small-scale outbursts observed in the light curve suggest naturally a pulsation analysis in each time interval between them to search for potential correlations between the NRP regime and the outbursts, as observed in \citet{hua09}. 

\begin{figure}
 \resizebox{\hsize}{!}{\includegraphics[angle=90]{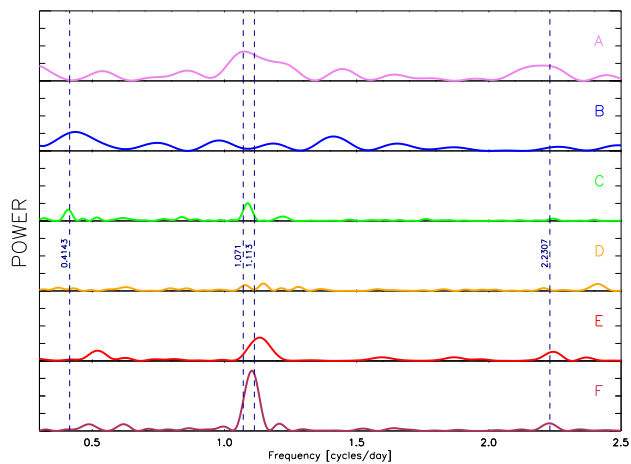}}
 \caption{CLEANEST power spectrum calculated for each piece (A) through
(F) in Fig.~\ref{lightcurve}, similarly colored, in order to evaluate the effect of
outbursts. The dashed lines indicate four of the most powerful
frequencies in the full light curve (Fig.~\ref{cleanest}).}
 \label{sburst}
 \end{figure}

A moving window frequency analysis was made to investigate the frequency multiplet apparently found (Figure~\ref{fig03}). We divide the photometric light curve into six regions (from A to F in  Fig.~\ref{lightcurve}) that is, between CJD = 3847, 3853, 3876, 3894, and 3905.  For each region, a CLEANEST spectrum was made  (cf. Fig.~\ref{sburst}). Although there is considerable loss of resolution because of the smaller time spans, some clear power variations can be seen. The dominant multiplet appears strong mainly in regions A, C, E, and F.  NRP power reaches a minimum in region D, after the medium-sized outburst observed at the end of region C. The strength variation in time of the multiplet can also be seen in Fig.~\ref{fig03}. It is possible that some of the pulsation energy was used to produce the outburst in HD\,171219. \citet{riv98} and \citet{hua09} found correlations between variability in NRP intensity and the Be phenomenon for two other Be stars. More extensive and systematic work is needed to clearly establish such a correlation.

\subsection{Ground-based data}
\subsubsection{Fourier analysis of spectral line profiles}

 \begin{figure*}
 \resizebox{\hsize}{!}{\includegraphics[angle=90]{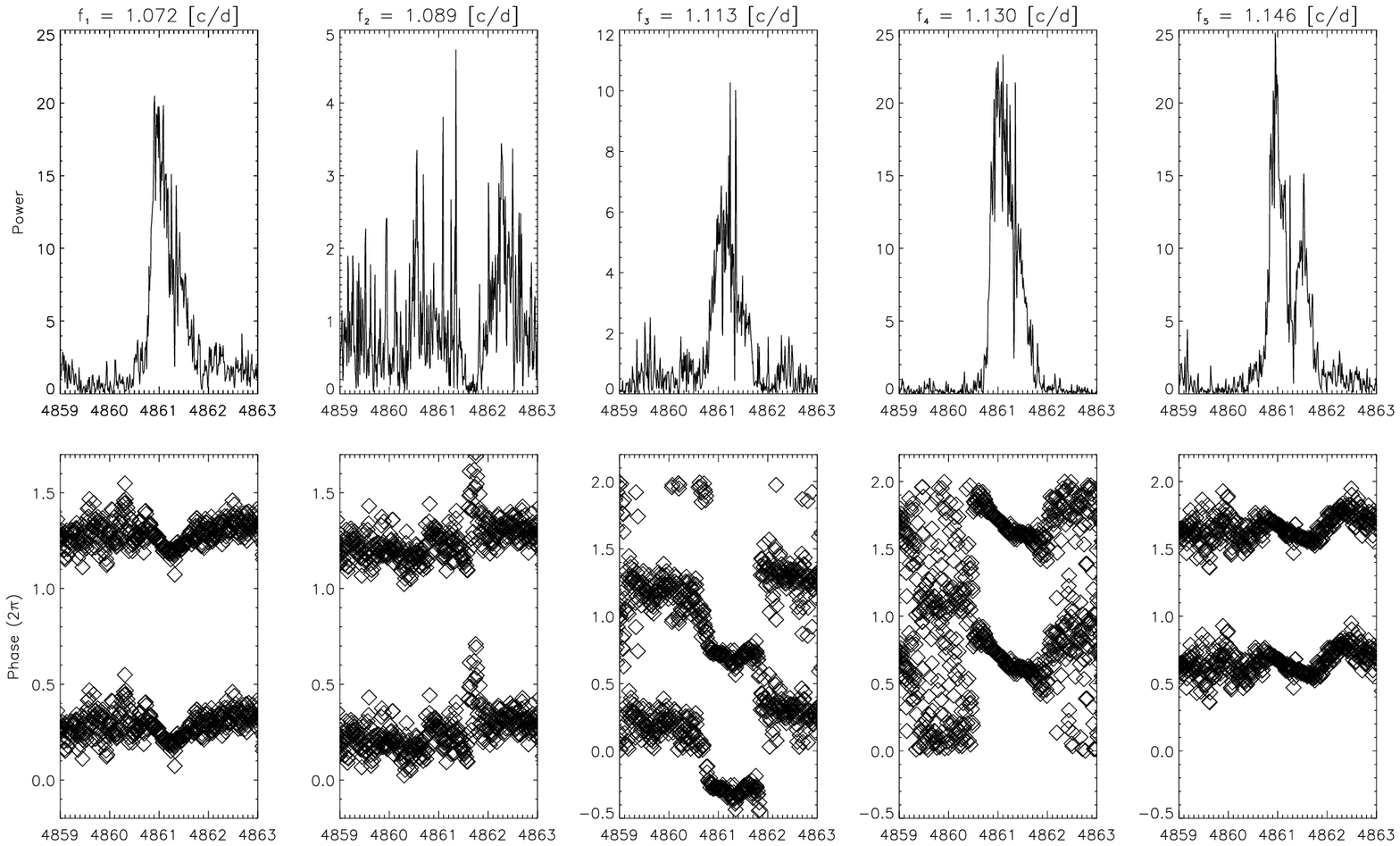}}
 \caption{Power (top panel) and phase (middle panel) variation of quintuplet frequencies along the mean HARPS and SOPHIE H$\beta$ line profile (Fig.~\ref{halpha}, bottom panel).}
 \label{fasequinto}
\end{figure*}  

We also applied the CLEANEST method to the time series of the spectroscopic data. Because of the scarcity of spectra, there was less information on NRP from line profile variations to be extracted \citep{nei12,jan00}. Nevertheless, peaks around the multiplet frequencies and their harmonics appear clearly. We fitted, by least squares, a Lorentzian function in the region of the multiplet, around $1.0$ c\,d$^{-1}$. The average half width at half-maximum (HWHM) found was $\Delta \nu = 0.026 \pm 0.009$ c\,d$^{-1}$ that we adopt as the frequency resolution of the spectroscopic data. Variations in those data can be visually inspected. We searched for phase variations in the H$\beta$ spectral series formed by HARPS and SOPHIE data for the main frequencies found with CoRoT (Table~\ref{table:3}).  

Coherent phase variations were found for some of the quintuplet frequencies. The results are presented in Fig.~\ref{fasequinto}.

As pointed out above, nonradial pulsations can contribute to trigger the Be phenomenon but \citet{riv13} found that for a majority of Be stars, an ejection velocity of about 50 km\,s$^{-1}$ is needed to form a Keplerian disk, even for very fast rotating stars. 

 \begin{figure}
 \resizebox{\hsize}{!}{\includegraphics[angle=90]{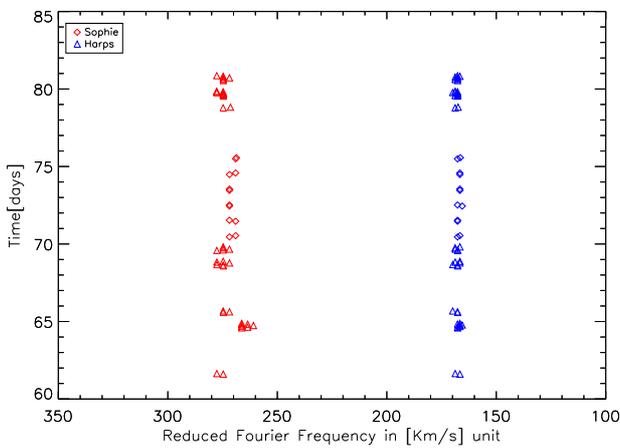}}
 \caption{Position of the first (red) and second (blue) minima of \ion{He}{i} $\lambda$ 4471 Fourier-transform profiles of HARPS and SOPHIE spectra.}
\label{velHeI}
\end{figure}

In order to look for possible velocity excursions caused by NRP, Fourier transform analysis has been applied to the \ion{He}{i} lines $\lambda$ 4026, 4144 and 4471 in each high-resolution spectrum of HD\,171219. The first minimum of the Fourier transform of line profiles can be reduced to velocity units so that it indicates the projected rotation velocity of the star~\citep{gra76}. However, the method works for symmetric, unblended lines, which is not the case of HD\,171219 (see Figs.~\ref{fit_complete} and~\ref{fit_parts}). Therefore, the V$\sin i$ will be underestimated, as can be seen in Fig.~\ref{velHeI}, which in fact shows real line width variations. These variations can be interpreted as being caused by a changing NRP velocity field super-imposed to the supposedly constant rotation velocity field. 

Variations in the line width first minimum of HD\,171219 occur nightly as well as during the same night with an amplitude of $\sim\,$ 15 km\,s$^{-1}$.  For a star seen almost equator-on, this is near the true amplitude of the velocity field and is not enough to trigger mass ejection episodes.  The same kind of results were obtained for the Be star $\zeta$\,Oph by \citet{jan00}. Indeed, such a velocity amplitude is compatible to predictions by \citet{dzi93}, who showed that g-modes of harmonic degree $\ell \leq 3$ are unstable for early to late B stars. They predicted velocity amplitudes caused by NRP between $\sim$ 10 and 30 km\,s$^{-1}$ for pulsation periods around $\sim 1$ day. Thus, the velocity excursions detected in HD\,171219 are probably insufficient to fully explain mass ejection episodes.

\subsubsection{NRP and line-profile variations}

\citet{tel97} showed that pulsation quantum numbers $\ell$ and $m$ can be estimated in a series of spectra from the blue-to-red phase variations of the observed pulsation frequency and its first harmonics, respectively. The method is especially valid for rapid rotators as is the case of HD\,171219. Again, our spectroscopic observations are limited, yielding a resolution in frequency of the order of just $0.026$ c\,d$^{-1}$. This resolution hinders a precise analysis of the CoRoT frequencies. Nevertheless, variations in spectroscopic data can be inspected visually. We thus searched for phase variations in the H$\beta$ spectral series formed by HARPS and SOPHIE data for the main frequencies found with CoRoT (Table~\ref{table:3}).

In Fig.~\ref{fasequinto} we plotted the amplitudes and phases of the variations of the multiplet frequencies as a function of the position in the H$\beta$ Balmer line profile. They were calculated by CLEANEST along the mean SOPHIE + HARPS observations. Two cycles of pulsations were plotted shifted by multiples of $\pi$ for clarity.  For each frequency in Fig.~\ref{fasequinto} we read off the blue-to-red phase differences within the region where the variational amplitudes are sufficiently strong. The phase differences may be underestimated because of the low variational power in the line wings.

Following \citet{tel97}, the blue-to-red phase variations of frequencies that showed clear variations along the H$\beta$ line were used to estimate the value of the NRP degree $\ell$. The value obtained for frequencies $f_3$ and $f_4$ yields $\ell = 1.7 \pm 0.2$. A harmonic degree $\ell = 2$ if rotationally split would produce a quintuplet (see Section 4.1.1) corresponding to a NRP order $m$ with values -2, -1, 0, +1 and +2. Some harmonic frequencies of the quintuplet seem to be present around $2.2$ c\,d$^{-1}$. In principle, pulsation orders $m$ could be determined from the phase variations across the H$\beta$ line \citep{tel97}. Unfortunately, the low intensity of the harmonics prevent their phases to be estimated.

\section{Conclusions} \label{sect:conclusions}

We have detected twenty periodicities in the CoRoT light curve of HD\,171219 compatible with nonradial pulsation g-modes (see Table~\ref{table:3}). Indeed, they are expected to be present in a B5e star. Frequencies $f_{1}$ to $f_{5}$ are identified as a quintuplet and some of their harmonics may be present. High-resolution spectroscopic data were obtained during the month preceding the CoRoT observations; a detailed analysis of spectral data points out to a quite rapidly rotating B5 III star, seen equator-on. The phase variations of fundamental tones along the H$\beta$ line are compatible with an azimuthal number $\ell = 2 \pm 1$. Four relatively small and a medium outbursts were observed in the light curve during the almost 80d time span of CoRoT observations. The frequency amplitudes vary neatly between outbursts. They pass through a minimum after the medium-sized outburst and this could be interpreted as pulsation power being temporarily transferred and consumed for pumping outbursts. It is also possible that the pulsations are less visible at the surface because they have been destabilized by the outbursts \citep[see][]{nei13}. Further detailed and extensive observations are still needed to establish a firm correlation between NRP regime and mass loss from Be stars.

\begin{acknowledgements}
We wish to thank the CoRoT team for the acquisition and reduction of the CoRoT data. This work was supported in Brazil by FAPESP (project 2008/57866), CNPq (projects 312581/2013-0 and 312738/2013-7) and CAPES (PNPD scholarship). EP acknowledges financial support from PRIN INAF 2014. MR acknowledges financial support from the  FP7-Space project SpaceInn. JCS acknowledges funding support from Spanish public funds for research under project ESP2015-65712-C5-5-R (MINECO/FEDER), and under Research Fellowship program ``Ram\'on y Cajal'' (MINECO/FEDER). This work has made use of the BeSS database, operated at LESIA, Observatoire de Meudon, France: http://basebe.obspm.fr.

\end{acknowledgements}

\bibliographystyle{aa}
\bibliography{andradenew}

\end{document}